\documentclass[prb,superscriptaddress,letterpaper,amsmath,floatfix,nofootinbib]{revtex4}
\usepackage{graphicx,psfrag}
\usepackage{amssymb}
\usepackage{citesort}
\usepackage{color}
\begin{document}

\title{\bf Coulomb Enhanced Superconducting Pair Correlations in the Frustrated Quarter-Filled Band.}
\author{Niladri Gomes}
\affiliation{Department of Physics, University of Arizona, Tucson, AZ 85721}
\author{W. Wasanthi De Silva}
\affiliation{Department of Physics \& Astronomy and HPC$^2$ Center for Computational Sciences, Mississippi State, MS 39762}
\author{Tirthankar Dutta}
\affiliation{Department of Physics, University of Arizona, Tucson, AZ 85721}
\author{R. Torsten Clay$^*$}
\affiliation{Department of Physics \& Astronomy and HPC$^2$ Center for Computational Sciences, Mississippi State, MS 39762}
\author{S. Mazumdar$^\dagger$,}
\affiliation{Department of Physics, University of Arizona, Tucson, AZ 85721}
\date{\today}
\begin{abstract}
{\bf A necessary condition for superconductivity (SC) driven by
  electron correlations is that electron-electron (e-e) interactions
  enhance superconducting pair-pair correlations, relative to the
  noninteracting limit.  We report high-precision numerical
  calculations of the ground state within the frustrated two-dimensional (2D)
  Hubbard Hamiltonian for a wide range of carrier concentration $\rho$
  ($0<\rho<1$) per site. We find that long range superconducting pair
  correlations are enhanced only for $\rho \simeq 0.5$.  At all other
  fillings e-e interactions mostly suppress pair correlations.  The
  enhancement of pair correlations is driven by the strong tendency to
  local singlet bond formation and spin gap (SG) in $\rho=0.5$, in
  lattices with quantum fluctuation \cite{Clay03a,Clay05a,Dayal11a}.
  We also report determinantal quantum Monte Carlo (DQMC) calculations
  that are in strong agreement with our ground state results.  Our
  work provides a key ingredient to the mechanism of SC in the 2D
  organic charge-transfer solids (CTS), and many other unconventional
  superconductors with frustrated crystal lattices and $\rho \simeq
  0.5$, while explaining the absence of SC in structurally related
  materials with substantially different $\rho$.  }
\end{abstract}
\maketitle

\let\thefootnote\relax\footnote{$^*$ Email address: {\bf r.t.clay@msstate.edu}}
\footnote{$^\dagger$ Email address: {\bf sumit@physics.arizona.edu}}

The possibility that e-e interactions can be the driving
force behind SC in correlated-electron systems has been intensely
investigated since the discovery of SC in the high T$_c$ cuprates. The
minimal requirements for a complete theory are, (i) the
superconducting pair correlations are enhanced by e-e interactions,
and (ii) pair correlations are long range.  For moderate to large e-e
interactions, pair correlations are perhaps best calculated
numerically, which however can be done only for finite
clusters. The simplest model incorporating e-e interactions
is usually assumed to be the Hubbard model, which in quite general form
can be written as
\begin{equation}
  H = -\sum_{\langle ij\rangle,\sigma}t_{ij} (c^\dagger_{i,\sigma}c_{j,\sigma}+H.c.) + U\sum_i n_{i,\uparrow}
  n_{i,\downarrow} + \frac{1}{2}\sum_{\langle i j\rangle} V_{ij} n_i n_j.
  \label{hamuv}
\end{equation}
All terms in Eq.~\ref{hamuv} have their standard meaning. The first
sum is the kinetic energy of noninteracting electrons within a 2D
tight-binding model with hopping matrix elements $t_{ij}$; $U$ and
$V_{ij}$ are onsite and nearest neighbor (n.n.) Coulomb interactions respectively.
Existing numerical calculations within Eq.~\ref{hamuv} on 2D lattices have failed to
find enhancement of pair-pair correlations relative to the
noninteracting model without making severe approximations
\cite{Scalapino12a}.

It has sometimes been surmised that correlated-electron SC might
evolve upon doping a spin-gapped semiconductor, as would occur in toy
models such as a 2D lattice consisting of weakly coupled even-leg
ladders \cite{Troyer96a,Arrigoni04a}.  Finding realistic 2D models
with SG and enhanced pair correlations however remains challenging.
In the present work we demonstrate from explicit numerical
calculations on frustrated 2D lattices enhanced pair correlations
evolving from a spin-gapped state at a carrier density $\rho\simeq
0.5$, far from the region most heavily investigated until now
($0.7<\rho<1.0$). We further point out the strong relevance of the
resulting theoretical picture to real materials, in particular the 2D
CTS superconductors, which were discovered earlier than the cuprates
\cite{Ishiguro} but are still not understood.

\begin{figure}[tb]

  \vspace*{2.0in}
  \begin{minipage}{3.0in}
    \centerline{\resizebox{3.0in}{!}{\includegraphics{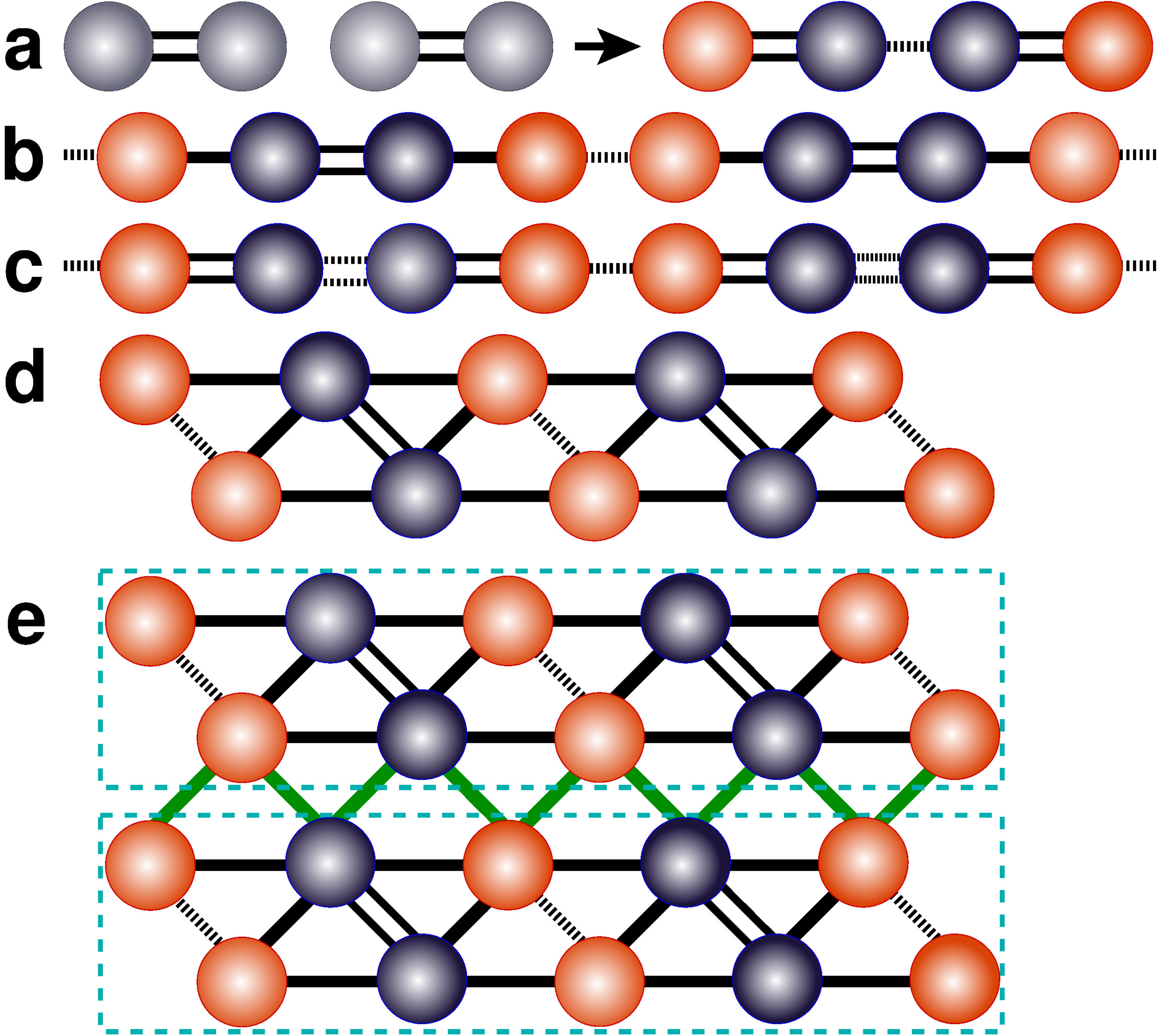}}}
  \end{minipage}
  \begin{minipage}{3.5in}
    \centerline{\resizebox{3.0in}{!}{\includegraphics{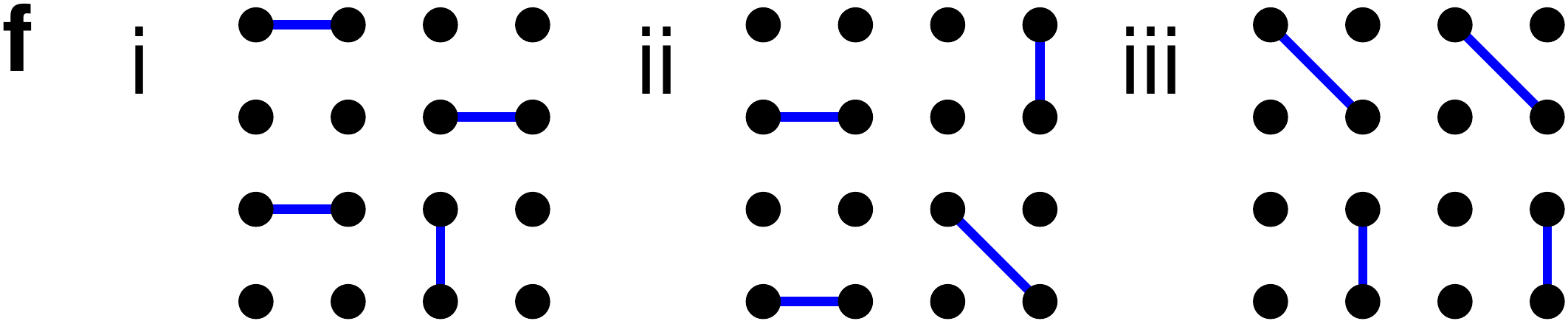}}}
    \vskip 0.2in
    \centerline{\resizebox{3.0in}{!}{\includegraphics{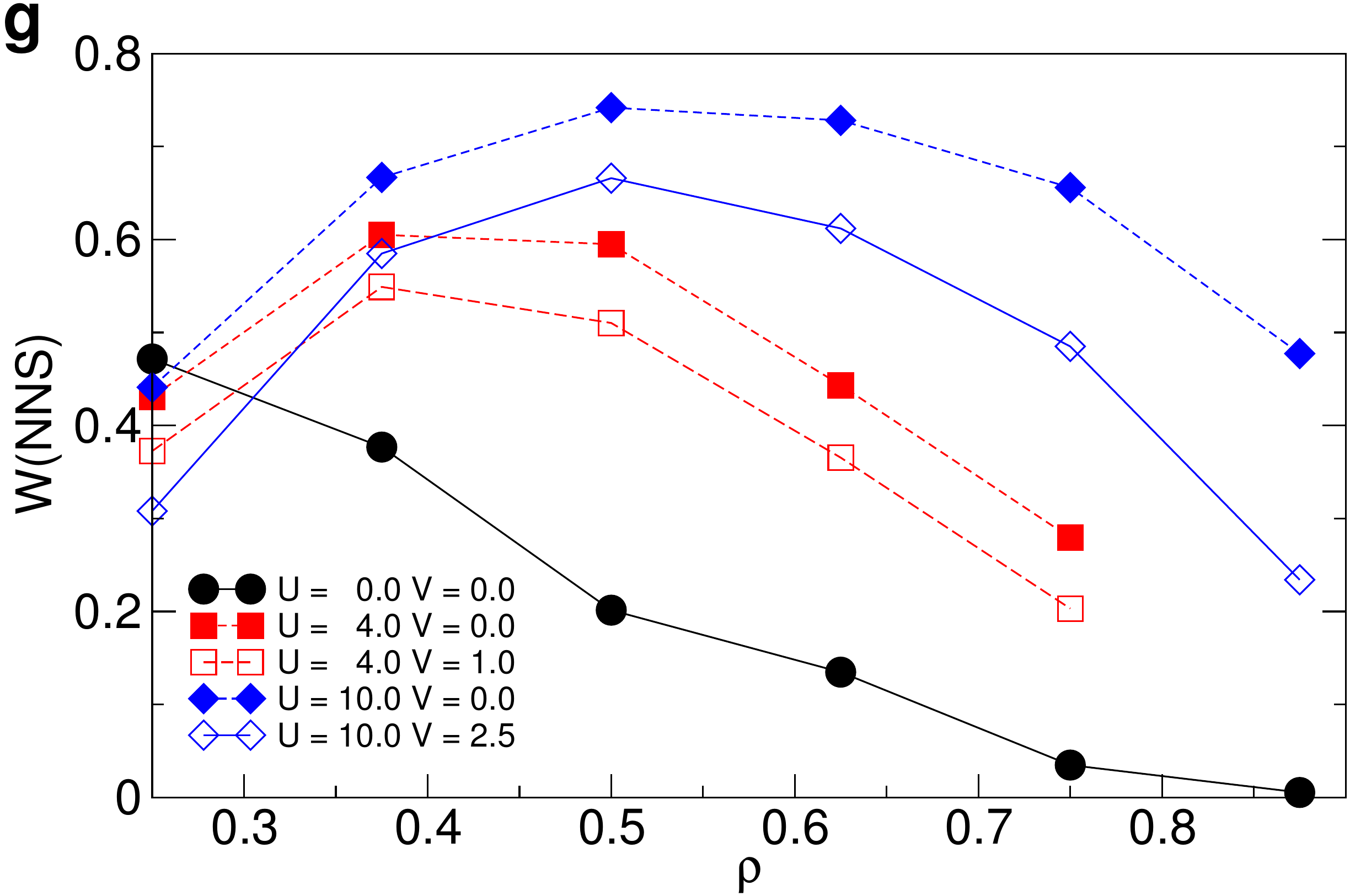}}}
  \end{minipage}
  \vspace*{1.0in}
  
  \caption{{\bf Effective e-e attraction in $\rho=0.5$.} (a)
    $\rho=0.5$ dimers with weak (left) and moderate (right) interdimer
    singlet bonding. Sites colored gray, blue and red have charges
    0.5, $>0.5$ and $<0.5$, respectively. (b) and (c) 2k$_F$ spin
    singlet states in the $\rho=0.5$ 1D chain, for small to
    intermediate $U$ and $V$ ($U\simeq4$, $V\simeq1$), and for
    intermediate to large $U$ and $V$ ($U \leq 10$, $V \leq 3$)
    respectively.  In both cases $V<V_c(U)$ \cite{Clay03a}.  (d) The
    PEC in the $\rho=0.5$ zigzag ladder \cite{Clay05a}. (e) The PEC in
    the anisotropic 2D triangular lattice \cite{Dayal11a}. The CO has
    pattern $\ldots$1100$\ldots$ in two directions and
    $\ldots$.1010$\ldots$ in the third direction, where `1' and `0'
    denote charge-rich and charge-poor sites.  The 2D PEC has the same
    charge structure as would be obtained from coupled zigzag
    ladders. Double, single and dotted bonds in (b) -- (e) denote
    bonds with decreasing strengths, with the double dotted bond
    weaker than a single bond but stronger than a single dotted bond.
    Differences in bond strengths result from nonzero e-p
    coupling. (f) Covalent VB diagrams with short bonds in
    $\rho=0.5$. (g) Exact total normalized weights of covalent VB
    diagrams with short bonds in the ground state wavefunction of
    different $\rho$ for the 4$\times$4 lattice, for $t_y=1$,
    $t_{x+y}=0.8$.}
  \label{singlets}
  
\end{figure}

There occurs an effective e-e attraction uniquely at $\rho=0.5$, in
lattices with strong quantum fluctuations, driven by
charge-spin-lattice coupling.  Consider the four-atom dimerized
``molecule'' of Fig.~\ref{singlets}(a), with two strong intradimer
bonds and one electron on each dimer. In the absence of the interdimer
bond, electron populations are equal on all sites. As the interdimer
electron hopping is switched on slowly, there is net migration of
charge to the two center atoms, due to the attractive
antiferromagnetic spin-coupling which leads to a singlet bond
\cite{Dayal11a}.  The charge migration is enhanced by the presence of
electron-phonon (e-p) interactions \cite{Clay03a,Dayal11a}. The
effective attraction is stronger than that near $\rho \sim 1$, where
the tendency to such charge migration is necessarily smaller, with the
neighboring sites already occupied.  The charge-ordering (CO) of
Fig.~\ref{singlets}(a) in the spin-singlet state persists in the
thermodynamic limit in one dimension (1D) $\rho=0.5$, where for
$V<V_c(U)$ e-e and e-p couplings act co-operatively \cite{Clay03a} to
give the spin-Peierls state with the 2k$_F$ bond modulations of
Fig.~\ref{singlets}(b) and (c).  Phase segregation is avoided, as the
key requirement for charge migration is a final state with vacancies
on both sides of the singlet bond (see Fig.~\ref{singlets}(a)).  The
spin-Peierls state at $\rho=0.5$ is a {\it paired-electron crystal}
(PEC), in which singlet-coupled n.n. singly occupied sites are
separated by pairs of vacancies.  Similar PECs occur in the $\rho=0.5$
zigzag ladder (Fig.~\ref{singlets}(d)) \cite{Clay05a} and in the 2D
anisotropic triangular lattice (Fig.~\ref{singlets}(e)) for
sufficiently large lattice frustration \cite{Dayal11a}. We have not
found the PEC \cite{Dayal11a} at any other $\rho$.  The exceptional
stability of the PEC at $\rho=0.5$ is due to its commensurate
structure.  The PEC has been experimentally observed in a number of
CTS \cite{Tamura06a,Drichko14a} (see Supplemental Information).

Based on a valence bond (VB) perspective that has some overlap with
Anderson's resonating valence bond \cite{Anderson87a} approach to the
nearly $\rho=1$ limit, we posit that SC is achieved in $\rho \simeq
0.5$ upon destabilization of the PEC, either due to increased
frustration or very weak doping.  The PEC wavefunction is dominated by
covalent VB diagrams with periodic arrangement of the n.n. singlet
bonds. Close to the PEC, the static CO and bond order are lost, but we
anticipate the wavefunction to continue to be dominated by VB diagrams
with singlet bonds between n.n. charge-rich sites, except that the
arrangement of the bonds is no longer periodic.  One such diagram is
shown in Fig.~\ref{singlets}(f)(i).  Within Eq.~\ref{hamuv}, pairs of
VB diagrams with only n.n. bonds are coupled through the diagrams with
next nearest neighbor (n.n.n.) bonds, as in
Figs.~\ref{singlets}(f)(ii) and (iii).  We collectively refer to
diagrams with only n.n. and n.n.n. bonds as those with short bonds.
There will be considerable pair tunneling in a wavefunction dominated
by VB diagrams with short bonds, and we will refer to such a
wavefunction as a paired electron liquid (PEL). The PEL would be most
stable near $\rho=0.5$, where the gain in kinetic energy due to pair
tunneling is the largest.

We consider an anisotropic triangular lattice with $t_{ij}=\{t_x$,
$t_y$, $t_{x+y}\}$.  $V_{ij}$ can similarly have three components.  We
express all quantities with dimensions of energy in units of $t_x$
($t_x=1$). The bulk of our calculations are for $t_y \simeq 1$, with
$t_{x+y}$ only slightly smaller. This is because broken symmetries
other than SC, viz., antiferromagnetism (AFM) and CO
dominate at weaker frustrations \cite{Dayal11a}.  We first calculate
the exact wavefunctions in the lowest total spin $S=0$ subspace for
all $\rho$ within the periodic 4$\times$4 triangular lattice.  In
Fig.~\ref{singlets}(g) we plot the total normalized contribution by
the covalent VB diagrams with short bonds to the exact wavefunction as
a function of $\rho$ for several Hubbard $U$ and $V$. For moderate to
large e-e interactions the maximum in this contribution occurs at
$\rho=0.5$, where the wavefunction consists predominantly of VB
diagrams with short bonds, indicating that VB diagrams with short
bonds as in Fig.~\ref{singlets}(f) dominate at $\rho=0.5$.

In the PEL the occurrence of local singlets is enhanced by
correlations and peaks for $\rho\simeq0.5$.  We anticipate Bose
condensation of singlet pairs within the PEL state within the
mechanism of SC proposed by Schafroth \cite{Schafroth55a}.  Without
e-p coupling in Eq.~\ref{hamuv} there is however no static SG and PEC
insulating state, which suggests that a complete theory of SC will
necessarily require explicit inclusions of both e-e and dynamic e-p
interactions.  As is however well established from studies of CDWs and
SDWs, the {\it tendency} to the dominant instability in models
containing both e-e and e-p interactions can be determined from
studying correlation functions of the electronic Hamiltonian alone
\cite{Hirsch84a}.  We have therefore performed calculations within
Eq.~\ref{hamuv} to determine if the dominance of VB diagrams with
short singlet bonds at $\rho\simeq0.5$ is accompanied by enhanced
long-range superconducting pair correlations.  The goal is to
demonstrate that the PEL is a distinct ground-state phase and is a
{\it precursor} to a correlated superconducting state.

We define the standard singlet pair-creation operator
\begin{equation}
  \Delta^\dagger_i = \sum_\nu g(\nu)\frac{1}{\sqrt{2}}(c^\dagger_{i,\uparrow}c^\dagger_{i+\vec{r_\nu},\downarrow}
  - c^\dagger_{i,\downarrow}c^\dagger_{i+\vec{r_\nu},\uparrow}),
\end{equation}
where $g(\nu)$ determine the pairing symmetry. We have calculated the
equal-time pair-pair correlations $P_{ij}=\langle \Delta^\dagger_i
\Delta_j\rangle$, for four different periodic anisotropic triangular
lattices, 4$\times$4, 6$\times$6, 10$\times$6 and 10$\times$10, with
the widest possible carrier densities $0 < \rho <1$, using three
different numerical techniques (see Methods). We have chosen lattices
that have a closed shell configuration at $\rho$ exactly 0.5, see also
Supplementary Information.  To facilitate comparison of multiple
lattices and to mitigate finite-size effects, we calculate the
distance dependent pair-pair correlations $P(r)$ ($r\equiv|\vec{r_i}
-\vec{r_j}|$) and show here the average long-range pair-pair
correlation $\bar{P} = N_P^{-1} \sum_{|\vec{r}|>2} P(r)$, where $N_P$
is the number of terms in the sum \cite{Huang01a}.

\begin{figure}[tb]

  \begingroup
  \sbox0{\includegraphics{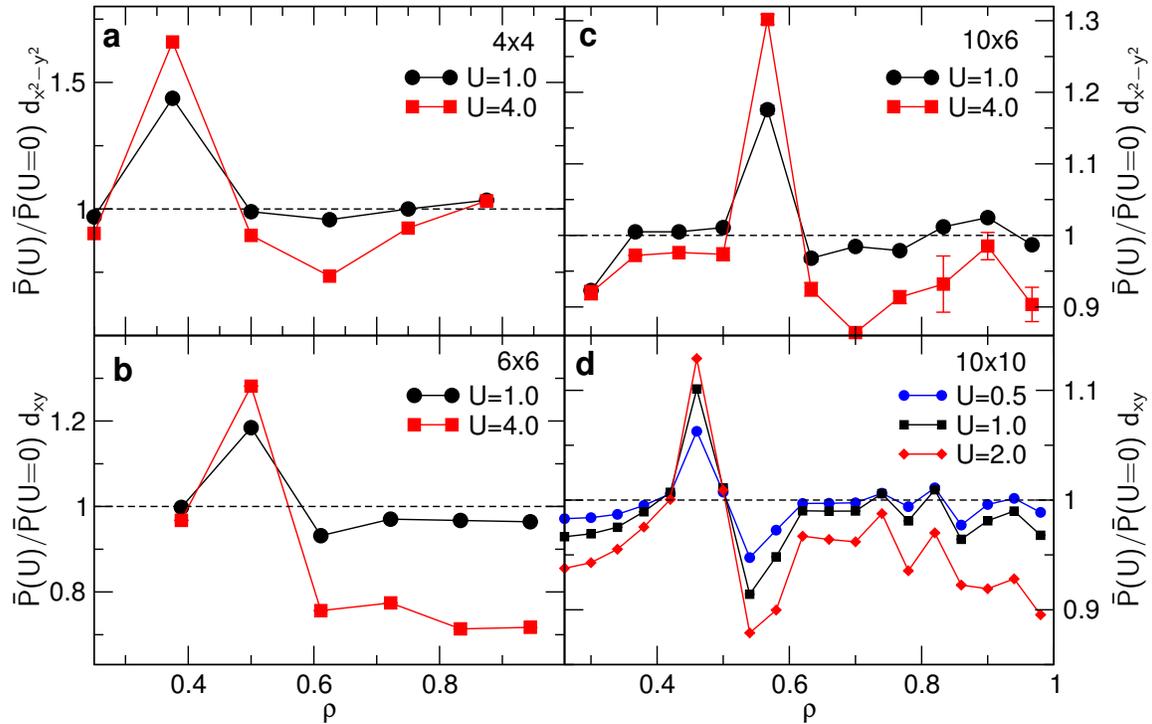}}%
  \includegraphics[clip,trim={.2\wd0} {.5\ht0} {.2\wd0} 0,width=6.0in]{Fig2}
  \endgroup

  \caption{{\bf Density dependence of dominant ground state pair-pair
      correlations.} Average long range pair-pair
    correlation $\bar{P}(U)$ normalized by its uncorrelated value
    $\bar{P}(U=0)$, for the (a) 4$\times$4, (b) 6$\times$6, (c)
    10$\times$6 and (d) 10$\times$10 anisotropic triangular lattices,
    for $t_y=0.9$ and $t_{x+y}=0.8$. The 4$\times$4 results are exact;
    6$\times$6 and 10$\times$6 results are obtained using the PIRG
    method; and 10$\times$10 by the CPMC
    method. $\bar{P}(U)/\bar{P}(U=0)>1$ for a single $\rho$ in each
    case, either for $\rho=0.5$ or for one of two closest carrier
    fillings with closed shell Fermi level occupancies at $U=0$.}
\label{pairing-fig}
\end{figure}

\begin{figure}[tb]

  \begingroup
  \sbox0{\includegraphics{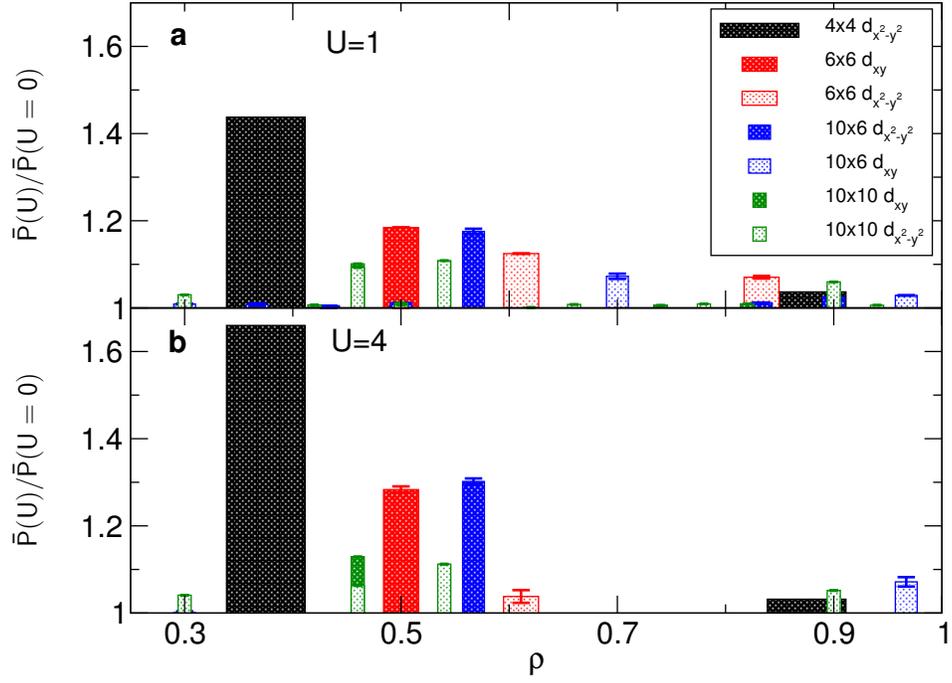}}%
  \includegraphics[clip,trim={.2\wd0} {.5\ht0} {.2\wd0} 0,width=5.0in]{Fig4}
  \endgroup

  \caption{{\bf Summary of $\rho$-dependence of ground state pair-pair
      correlation enhancement.}  Average long range pair-pair
    correlation $\bar{P}(U)$ normalized by its uncorrelated value
    $\bar{P}(U=0)$, for (a) $U=1$ and (b) $U=4$ ($U=2$ for the
    10$\times$10 lattice).  All results with
    $\bar{P}(U)/\bar{P}(U=0)>1$ are included, for both $d_{x^2-y^2}$
    and $d_{xy}$ pairing symmetries. The dominant pairing symmetry for
    each lattice is indicated with darker shading. The width of each
    bar is $1/N$, where $N$ is the number of lattice sites.}
  \label{pairing-summary}
  
\end{figure}

We have found $d_{x^2-y^2}$ and $d_{xy}$ symmetries to dominate over
$s$-wave symmetries in our calculations.  Further, for each lattice
only one of the two $d$-wave channels is relevant; $d_{x^2-y^2}$ for
4$\times$4 and 10$\times$6, and $d_{xy}$ for 6$\times$6 and
10$\times$10 (see Supplementary Information).  The origin of this
lattice dependence is currently not understood; note, however, that
the distinction between d$_{x^2-y^2}$ and d$_{xy}$ symmetries is to a
large extent semantic in the strongly frustrated regime we
investigate. Furthermore, it is possible that the actual pairing
symmetry is a superposition of $d_{x^2-y^2}$ and $d_{xy}$. We have not
attempted to find this superposition. Rather, for each lattice and
$\rho$ we have calculated the dominant symmetry $\bar{P}$ as a
function of $U$. Plots of $\bar{P}$ versus $U$ for the different
lattices and $\rho$ are given in the Supplementary Information.  The
complete results, summarized in
Figs.~\ref{pairing-fig} and \ref{pairing-summary}, are remarkable:
Coulomb interactions enhance superconducting pair correlations only in
a narrow density range close to $\rho=0.5$. For each lattice
$\bar{P}(U)/\bar{P}(U=0)>1$ for a single $\rho$ that is either exactly
0.5 or one of two closest carrier fillings with closed shell Fermi
level occupancy at $U=0$.  Pair correlations are suppressed by $U$ at
all other $\rho$, including the region $0.7<\rho<1$ that has been
extensively investigated in the context of cuprate SC
\cite{Scalapino12a}.  In three of four lattices in
Fig.~\ref{pairing-fig} enhancement of $\bar{P}(U)$ occurs for $\rho$
slightly away from 0.5.  The magnitude of pair correlations depend on
both the pair binding energy and the kinetic energy to be gained from
pair delocalization; in finite lattices both quantities depend
strongly on the details of the one-electron energy spectrum. We show
in the Supplementary Information that for each of the four lattices
the $\rho$ at which enhanced $\bar{P}(U)$ occurs can be predicted from
the known one-electron levels.  Importantly, the deviation from 0.5 of
the $\rho$ at which $\bar{P}(U)$ is enhanced (excluding the 6$\times$6
lattice where this deviation is zero) decreases monotonically with
size.

Having nonzero $V_{ij}$ affects lattice frustration minimally when all
three components, $V_x$, $V_y$ and $V_{x+y}$ are
nonzero. Pair-correlations for $V_x=V_y=V_{x+y}$ could be calculated
only for the 4$\times$4 lattice, where the behavior of the pair
correlations is qualitatively similar to $V_{ij}=0$, although the
magnitude of the enhancement is smaller.  We have found that when
$V_x=V_y$, $V_{x+y}=0$, $d_{xy}$ pair correlations are enhanced
uniquely for $\rho \simeq 0.5$.  Similarly, $V_{x+y} \neq 0$, and any
one of $V_x$, $V_y$ nonzero enhances (suppresses) $d_{x^2-y^2}$
($d_{xy}$). Overall, there is a broad parameter region over which the
pair correlations remain enhanced at the same $\rho$ where enhancement
is found for $U \neq 0$. (see Supplementary Information).

\begin{figure}[tb]

  \begingroup
  \sbox0{\includegraphics{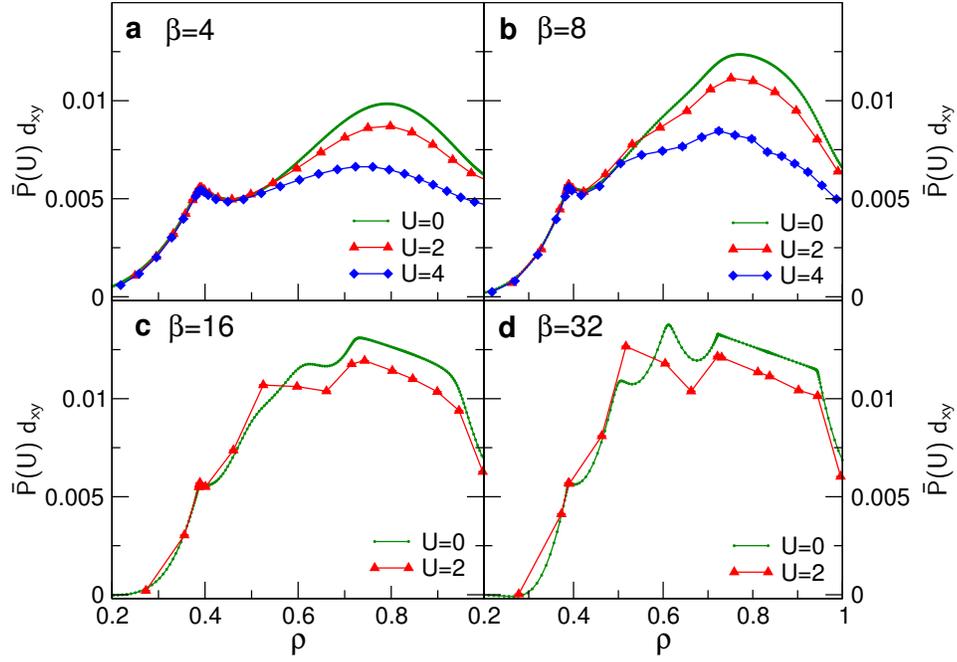}}%
  \includegraphics[clip,trim={.2\wd0} {.5\ht0} {.2\wd0} 0,width=5.0in]{Fig3}
  \endgroup

  \caption{{\bf Temperature dependence of pair-pair correlations, as
      calculated using DQMC.} $\bar{P}(U)$ for $d_{xy}$ pairing as a
    function of $\rho$ and inverse temperature $\beta$ for the
    6$\times6$ lattice with $t_x=$1, $t_y=$0.9, and $t_{x+y}$=0.8,
    calculated using DQMC.  Note the gradual enhancement of
    $\bar{P}(U)$ for $\rho \simeq 0.5$ with increasing $\beta$,
    beginning from $\beta=8$.  $\bar{P}(U)$ is suppressed by $U$ at
    all other $\rho$.}
  \label{QMC}
  
\end{figure}

The numerical ground state results are further confirmed by DQMC
calculations.  The sign problem is severe for large $U$, but up to
$U=2$ the results are reliable even for the largest $\beta$
($\beta=t_x/k_BT$) we have investigated (see Methods and Supplementary
Information).  Fig.~\ref{QMC} shows that with increasing $\beta$ there
occurs progressive enhancement of $\bar{P}(U)$ with increasing $U$,
uniquely at $\rho \simeq 0.5$.  As in Fig.~\ref{pairing-fig}, $U$
suppresses pair correlations at all other $\rho$ at large $\beta$. The
excellent agreement between $\bar{P}(U)$ obtained from Path Integral
Renormalization Group (PIRG) and DQMC indicates that while the DQMC
calculations could be performed at the smallest $T$ only for $U \leq
2$, enhanced pair correlations should be expected at even larger $U$.
Fig.~\ref{pairing-summary} summarizes the enhancement of pairing as a
function of $\rho$ for all lattices, including in the nondominant
channels.  As seen in Fig.~\ref{pairing-summary}, the dominant pairing
symmetry is enhanced only for $\rho \simeq 0.5$. Pairing in the
non-dominant channels is enhanced weakly for small $U\approx1$ for
some $\rho$, but are weakened further as $U$ is increased.

Our computational results have direct implication for the mechanism of
SC in the quasi-2D CTS.
Typical
superconducting CTS are the families (BEDT-TTF)$_2$X and
Z[Pd(dmit)$_2$]$_2$.
The number of holes (electrons) $\rho$ per BEDT-TTF cation
(Pd(dmit)$_2$ anion) is thus 0.5. The same stoichiometry is true for
all superconducting (but not merely conducting) CTS with different
organic molecular components.  Strong e-e interaction and frustrated
anisotropic triangular lattices also are common features
\cite{Kanoda11a,Powell11a}. Based on the crystal structures
(BEDT-TTF)$_2$X are referred to as $\alpha$, $\beta$, $\theta$,
$\kappa$ etc.  The $\kappa$-family has been investigated the most
intensively, because of the proximity of magnetic phases to SC in
these \cite{Kanoda11a,Powell11a}. In this family, X=Cu[N(CN)$_2$]Cl
($\kappa$-Cl) is AFM at ambient pressure, X=Cu$_2$(CN)$_3$
($\kappa$-CN) is a quantum spin liquid (QSL), and X=Cu[N(CN)$_2$]Br
($\kappa$-Br) and X=Cu(NCS)$_2$ ($\kappa$-NCS) are superconductors at
ambient pressure \cite{Kanoda11a}. SC is also observed in $\kappa$-Cl
and $\kappa$-CN under pressure \cite{Kanoda11a}. Pseudogap behavior
has been observed in the ``metallic'' $\kappa$-Br and $\kappa$-NCS for
T $<$ T$_{PG} \sim$ 50K from NMR
\cite{Mayaffre94a,deSoto95a,Kawamoto95a,Itaya09a,Kobayashi14a}, STM
\cite{Arai00a},
precision lattice expansion \cite{Muller02a}, and magnetic torque
measurements \cite{Tsuchiya12a}.

Structurally, in the $\kappa$-family the BEDT-TTF molecules are
strongly coupled as dimers, which form an anisotropic triangular
lattice. AFM in $\kappa$-Cl and QSL behavior in $\kappa$-CN are then
easily explained if the dimers and not the individual molecules are
considered as effective sites. This gives an effective $\rho=1$
Hubbard model that will yield the AFM (QSL) for weak (strong)
frustration \cite{Kanoda11a,Powell11a}. Precise numerical calculations
have however demonstrated the absence of SC within the $\rho=1$
Hubbard model for any frustration
\cite{Clay08a,Tocchio09a,Dayal12a}. Our present work is able to
explain both the magnetic behavior and SC: in the localized insulating
phase the dimerized $\rho=0.5$ and the effective $\rho=1$ model give
same behavior \cite{Dayal11a}, but in the pressure-induced delocalized
phase larger interdimer hopping leads to breakdown of the effective
picture and quantum critical transition from the AFM to the PEL, which
is superconducting once pair coherence is reached. Strong support for
this viewpoint is obtained from the recent observation of the PEC in
$\kappa$-Hg(SCN)$_2$Cl \cite{Drichko14a}. Further, our proposed
mechanism gives for the first time a cogent explanation of the
pseudogap in $\kappa$-Br and $\kappa$-NCS. The effective
$\rho=1$ model fails to explain the T $<$ T$_{PG}$ NMR behavior
\cite{Yusuf07a}. Within our theory at $T_{PG}$ there occurs the
quantum critical transition to the PEL, while phase coherence and SC
are reached only at T$_c$. Diamagnetism observed from Nernst
coefficient measurements at T $>$ T$_c$ \cite{Nam07a} supports this
picture of preformed pairs.
The close proximity of the PEL to the PEC 
 explains the density wave-like 
behavior in this phase noted by M\"uller et al. \cite{Muller02a}.
In Fig.~\ref{QMC}, enhanced pair
correlation at $\rho=0.5$ begins to appear already at $\beta=8$; with
average $|t| \sim 0.1$ eV, we see that T$_{PG}$ can be as high as
$\sim 100$ K.

In other CTS, the semiconducting state proximate to SC exhibits CO.
Pressure-induced SC from CO states is seen in $\alpha$- and
$\theta$-(BEDT-TTF) compounds \cite{Ishiguro},
EtMe$_3$P[Pd(dmit)$_2$]$_2$ \cite{Tamura06a} and
$\beta$-(meso-DMBEDT-TTF)$_2$X (X = PF$_6$ and AsF$_6$)
\cite{Shikama12a}.  The horizontal stripe CO below the SG transitions
in the $\alpha$- and $\theta$-(BEDT-TTF) \cite{Mori98b,Ivek10a}, the
so-called valence bond crystal order in EtMe$_3$P[Pd(dmit)$_2$]$_2$
\cite{Tamura06a} and the checkerboard CO in
$\beta$-(meso-DMBEDT-TTF)$_2$X (X = PF$_6$ and AsF$_6$)
\cite{Shikama12a} all have the same pattern as the PEC in
Fig.~\ref{singlets}(e) (see reference \onlinecite{Clay02a} and
Supplementary Information).  The bandwidth-driven quantum critical
transition is now directly from the PEC to SC.  The strong role of
phonons in SC seen experimentally \cite{Girlando02a} is expected, as
it is the co-operative effect between the e-e and e-p interactions
\cite{Clay03a,Dayal11a} that drives the transitions to the PEC and
PEL.

Two of us have recently pointed out the unusual abundance
\cite{Mazumdar14a} of seemingly unrelated correlated-electron
materials that are superconducting at carrier concentration $\rho
\simeq 0.5$. In all cases the superconductor belongs to a family of
materials with varying $\rho$, and no SC is observed for $\rho$
substantially different from 0.5 (as is also true for the CTS).  While
the experimental systems should be investigated individually, it is
conceivable that the shared features of $\rho=0.5$, lattice
frustration and strong e-e interaction point to a new paradigm for
correlated-electron SC.
\vskip 1pc

\noindent{\bf Acknowledgements.} This work was supported by the US
Department of Energy grant DE-FG02-06ER46315 (De Silva, Dutta, and
Clay) and by NSF-CHE-1151475 (Gomes and Mazumdar). Part of the
calculations were performed using resources of the National Energy
Research Scientific Computing Center (NERSC), which is supported by
the Office of Science of the US Department of Energy under Contract
No. DE-AC02-05CH11231.

\vskip 1pc
\noindent{\bf Methods.} Calculations used four different methods: i)
         {\it Exact diagonalization using the valence bond basis:}
         This exact method was used to calculate ground-state
         quantities for the 4$\times$4 lattice; ii) {\it Path Integral
           Renormalization Group (PIRG):} PIRG was used to calculate
         ground-state pair-pair correlation functions for the
         6$\times$6 and 10$\times$6 lattices; iii) {\it Constrained
           Path Monte Carlo (CPMC):} CPMC with a free-electron trial
         wavefunction was used to calculate pair-pair correlation
         functions for the 10$\times$10 lattice; iv) {\it
           Determinantal Quantum Monte Carlo (DQMC):} DQMC was used to
         calculate pair-pair correlation functions at finite
         temperature for the 6$\times$6, 10$\times$6, and 10$\times$10
         lattices.  Further details of the methods are given in the
         Supplementary Information.
\vskip 1pc

\noindent{\bf Author Contributions.} N. Gomes, W. Wasanthi De Silva,
and R. T. Clay performed PIRG, CPMC, and DQMC calculations. T. Dutta
performed the VB calculations. R. T. Clay and S. Mazumdar conceived of
the overall project and wrote the manusctript.

\vskip 1pc
\noindent{\bf Competing financial interests.}
The authors declare no competing financial interests.

\newpage

\clearpage

\clearpage

\clearpage

\end{document}